\documentclass[10pt, journal]{IEEEtran}%confermMence
\IEEEoverridecommandlockouts
\usepackage{cite}
\usepackage{amsmath,amssymb,amsfonts}
\usepackage{bm}
\usepackage{graphicx}
\usepackage{textcomp}
\usepackage{xcolor}
\usepackage[center]{caption}
\usepackage{amsthm}
\usepackage{bbm}
\usepackage{subfig}

\makeatletter
\newif\if@restonecol
\makeatother

\usepackage[linesnumbered,ruled,vlined]{algorithm2e}%[ruled,vlined]{
\usepackage{algpseudocode}
\usepackage{amsmath}
\def\BibTeX{{\rm B\kern-.05em{\sc i\kern-.025em b}\kern-.08em
    T\kern-.1667em\lower.7ex\hbox{E}\kern-.125emX}}
\begin{document}

\title{Empowering the Edge Intelligence by Air-Ground Integrated Federated Learning}%UAVs-Enabled Federated Learning in Air-Ground Integrated Networks}
%
%\author{Chao~Dong,
%        Yun~Shen,
%        Yuben~Qu,
%        Qihui~Wu,
%        Fan~Wu,
%        and~Guihai~Chen
%%\thanks{Copyright (c) 2015 IEEE. Personal use of this material is permitted. However, permission to use this material for any other purposes must be obtained from the IEEE by sending a request to pubs-permissions@ieee.org.}
%\thanks{C. Dong, Y. Shen, and Q. Wu are with the Key Laboratory of Dynamic Cognitive System of Electromagnetic Spectrum Space, Nanjing University of Aeronautics and Astronautics, China.}
%\thanks{Y. Qu (\textit{Corresponding author}), F. Wu, and G. Chen are with the Department of Computer Science and Engineering, Shanghai Jiao Tong University, China.}
%}

\author{Yuben~Qu,
        Chao~Dong,
        Jianchao~Zheng,
        Haipeng~Dai,
        Fan~Wu,
        Song~Guo,
        and~Alagan~Anpalagan
%\thanks{Copyright (c) 2015 IEEE. Personal use of this material is permitted. However, permission to use this material for any other purposes must be obtained from the IEEE by sending a request to pubs-permissions@ieee.org.}
\thanks{Y. Qu is with the Key Laboratory of Dynamic Cognitive System of Electromagnetic Spectrum Space, Nanjing University of Aeronautics and Astronautics, China, and also with the Department of Computer Science and Engineering, Shanghai Jiao Tong University, China (E-mail: quyuben@nuaa.edu.cn).}
\thanks{C. Dong (\textit{Corresponding author}) is with the Key Laboratory of Dynamic Cognitive System of Electromagnetic Spectrum Space, Nanjing University of Aeronautics and Astronautics, China (E-mail: dch@nuaa.edu.cn).}
\thanks{J. Zheng is with National Innovation Institute of Defense Technology, China (E-mail: longxingren.zjc.s@163.com).}
\thanks{F. Wu is with the Department of Computer Science and Engineering, Shanghai Jiao Tong University, China (E-mail: fwu@cs.sjtu.edu.cn).}
\thanks{S. Guo is with The Hong Kong Polytechnic University Shenzhen Research Institute and Department of Computing, The Hong Kong Polytechnic University, Hong Kong (E-mail: cssongguo@comp.polyu.edu.hk).}
\thanks{A. Anpalagan is with the Department of Electrical and Computer Engineering, Ryerson University, Canada (E-mail: alagan@ee.ryerson.ca).}
}

\maketitle

\begin{abstract}
Ubiquitous intelligence has been widely recognized as a critical vision of the future sixth generation (6G) networks, which implies the intelligence over the whole network from the core to the edge including end devices. Nevertheless, fulfilling such vision, particularly the intelligence at the edge, is extremely challenging, due to the limited resources of edge devices as well as the ubiquitous coverage envisioned by 6G. To empower the edge intelligence, in this article, we propose a framework called AGIFL (Air-Ground Integrated Federated Learning), which organically integrates air-ground integrated networks and federated learning (FL). In the AGIFL, leveraging the flexible on-demand 3D deployment of aerial nodes such as unmanned aerial vehicles (UAVs), all the nodes can collaboratively train an effective learning model by FL. We also conduct a case study to evaluate the effect of two different deployment schemes of the UAV over the learning and network performance. Last but not the least, we highlight several technical challenges and future research directions in the AGIFL.
\end{abstract}

%\begin{IEEEkeywords}
%edge computing
%\end{IEEEkeywords}
\section{Introduction}\label{Sec1}
%½éÉÜ6G£¬¶Ô6G×öÒ»¸ö¼òµ¥µÄÃèÊö(Í»³öÆä·ºÔÚÖÇÄÜÔ¸¾°)
Although the fifth generation (5G) networks are being deployed and about to be commercially available in 2020, it is predicted that the upcoming 5G networks may still not be able to meet future rapidly growing traffic demands. Accordingly, the beyond 5G (B5G) networks, or say sixth generation (6G), have already received great attention by both the academia and industry around the world. It is envisioned that an essential difference between 6G networks and previous generation networks lies in the revolution of ubiquitous intelligence realization, to enable colorful artificial intelligence (AI) services from the network core to the network edge including the end devices \cite{AI6G}. Nevertheless, realizing such ubiquitous intelligence of 6G is extremely challenging.

%Ö¸³ö±ßÔµÖÇÄܺͿյØÒ»ÌåÍøÂçÊÇÆäºËÐÄDZÔÚÇý¶¯¼¼ÊõÖ®Ò»£¬Ä¿Ç°¹ØÓÚ6G±ßÔµÖÇÄܵÄһЩÑо¿½éÉÜ£¬µ«ÈçºÎʵÏÖ±ßÔµÖÇÄÜÉÐÐèÑо¿£¬ÓÈÆäÊÇÔÚ¿ÕÌìµØº£Ò»ÌåÍøÂç±³¾°ÏÂ
Recently, edge intelligence (EI), where intelligence is pushed to the network edge by running AI algorithms on edge devices, emerges as a promising key enabler for 6G to fulfill the vision of ubiquitous intelligence \cite{V6G}. EI also caters to the trend of most of the big data originated from the center cloud to the network edge, with the proliferation of massive Internet-of-Things (IoTs) devices combined with various mobile applications. However, EI is still in its infancy, as it is indeed hard to implement distributed learning based on limited datasets across a huge number of heterogenous resource-constrained devices, especially in the context of ubiquitous coverage by integrating air networks, space networks, and underwater networks into 6G besides the conventional terrestrial networks.

%ÎÒÃÇÌá³öÁË¿ÕµØÒ»ÌåÁªºÏѧϰµÄ¼Ü¹¹×÷ΪʵÏÖ6G ·ºÔÚÖÇÄܵÄÒ»ÖÖÖØÒª·½Ê½
In this work, motivated by the recent advance of an emerging distributed machine learning (ML) methodology, \textit{i.e.}, federated learning (FL) \cite{CELD}, we propose a novel framework of \underline{A}ir-\underline{G}round \underline{I}ntegrated \underline{F}ederated \underline{L}earning (AGIFL), to boost the urgently needed EI in 6G. FL is a distributed learning architecture that enables multiple resource-constrained end devices to collaboratively train an effective learning model in a federated manner. Our proposed AGIFL is a marriage of FL and air-ground integrated networks (AGINs), where AGINs flexibly deploying unmanned aerial vehicles (UAVs), balloons, and airships with flying base stations are a critical part of 6G to support near-instant and seamless super-connectivity.

There exist limited few studies that investigate how to utilize FL to boost the intelligence for UAV networks in existing literature. Zeng \textit{et al.} \cite{FLS} propose a novel framework to enable FL within a swarm of UAVs, which is the first work considering how to implement FL for the UAV swarm. Shiri \textit{et al.} \cite{CEMU} study the online path control problem of massive UAVs by FL and mean field game (MFG) theory, where each UAV periodically exchanges the Hamilton-Jacobi-Bellman (HJB) neural network (NN) and Fokker-Planck-Kolmogorov (FPK) NN model parameters in MFG with other UAVs in a federated manner. Lim \textit{et al.} \cite{TFLU} propose an FL-based sensing and collaborative learning scheme for UAVs, where UAVs collect data and participate in the collaborative model training for Internet of Vehicles (IoV) applications. Brik \textit{et al.} \cite{FLUE} discuss several potential applications of federated deep learning (FDL) in UAVs-enabled wireless networks, and highlight the key challenges, open issues, and promising future research topics therein. Nevertheless, they mainly consider how to empower the aerial-part intelligence, where the FL model training is restricted in the aerial-part. In contrast, we focus on how to bring the edge intelligence into the whole AGINs, by proposing a comprehensive AGIFL framework where the FL model training exists widely in the AGINs, and the aerial-to-aerial FL in \cite{FLS,CEMU,TFLU,FLUE} can be actually viewed as a special case of the proposed AGIFL.

%This article aims to provide a preliminary attempt to realize the ubiquitous intelligence. The rest of this article is organized in the follows. We first introduce the basic concept of FL with its potential for 6G networks in Section~II. We then present the proposed AGIFL framework in Section~III and its different forms in Section~IV. In Section~V, we analyze the main technical challenges of the AGIFL, while we conduct a case study of AGIFL to evaluate its corresponding performance in Section~VI. Section~VII concludes the paper.

\section{FL and Its Potential for 6G Networks}\label{Sec2}
In this section, we first provide an overview of FL through an example of federated deep learning, and then discuss its potential for future 6G networks.
\begin{figure*}[t]
%\vspace{-0.3cm}
\centering
\includegraphics[scale = 0.375]{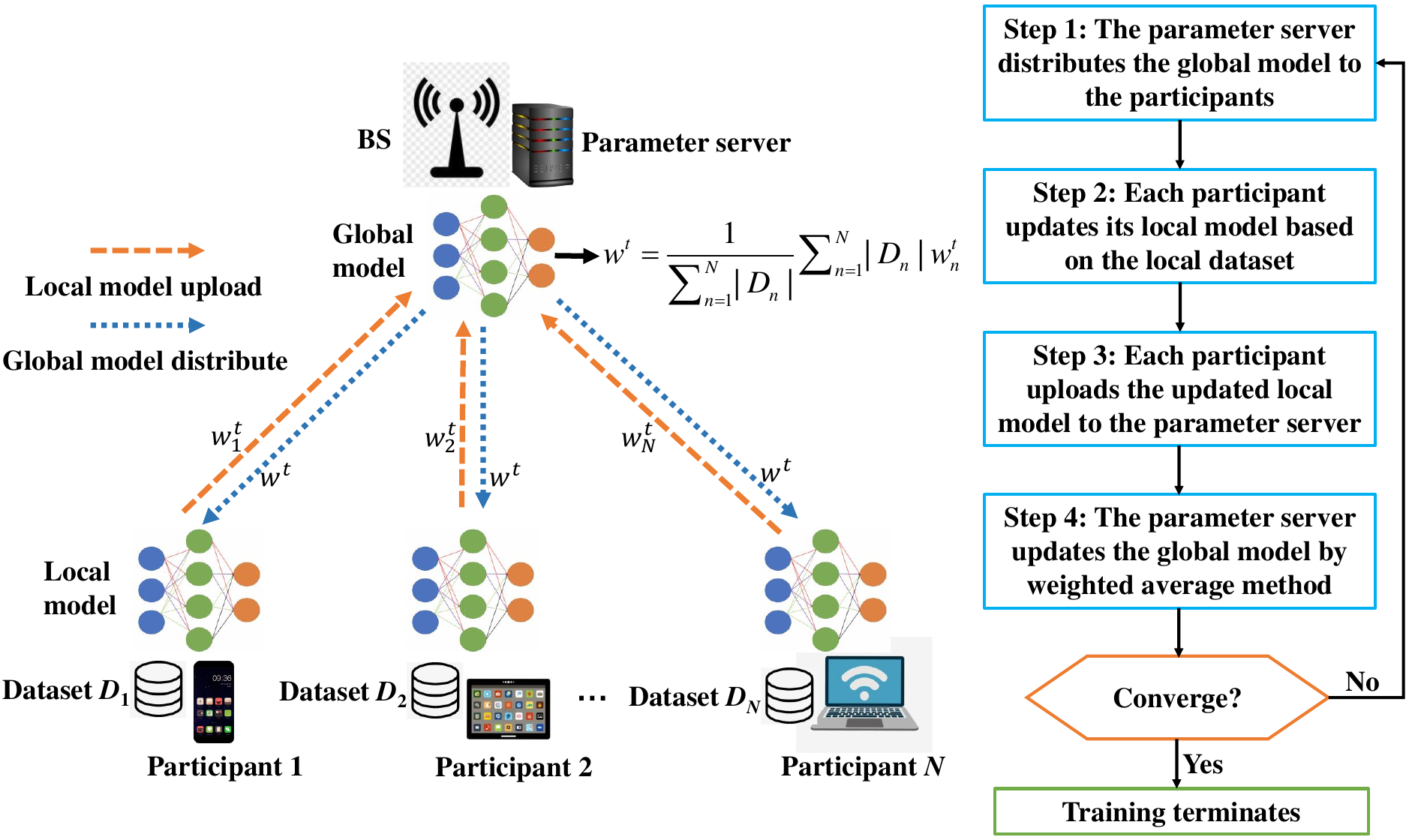}
\caption{An illustration of FedAvg \cite{CELD}.}
\label{FL_Illustration}
\vspace{-0.2cm}
\end{figure*}
\subsection{Overview of FL}
FL is an emerging distributed architecture which enables multiple end devices to collaboratively train a learning model in a federated way. The main advantage of FL lies in that it can utilize the limited on-device processing power and private local training data to distributively perform the model training, without allowing the data leave from its owner. FL was originally introduced by Google \cite{CELD}, which aims to train a clustering or classification model with training data such as images, videos or texts distributed over a large number of end devices, while reducing network overhead and alleviating privacy concerns. Although the initial FL refers to the federated deep learning framework, it can be actually used to train a FL model for many other advanced machine learning algorithms, \textit{e.g.}, federated deep reinforcement learning.

%\vspace{-0.35cm}

Next, we introduce the basic idea of FL through briefly describing the celebrated FedAvg scheme \cite{CELD}. As illustrated in Fig.~\ref{FL_Illustration}, there are generally two main entities in FL: the parameter server and multiple training participants who are data owners. With the help of the parameter server, FedAvg enables these participants to collaboratively train a shared learning model, while keeping all the local training data at each participant. First, the parameter server distributes the \textit{global model} to all participants (an initial global model is generated by the parameter server in the very beginning). Secondly, based on the latest received global model, each participant employs its own training data to update a \textit{local model}. Thirdly, each participant uploads the updated local model parameters (\textit{i.e.}, model weights) only to the parameter server. Lastly, the parameter server will aggregate all received model weights by the weighted average method to update the global model. If the converge condition is satisfied, the FL training terminates; otherwise, the above iteration continues. Note that the traditional ML approaches employ the entire training dataset consisting of the data at each user/participant, to train a global learning model in a centralized manner.

%\begin{algorithm}[t]
%  \footnotesize
%  \caption{FedAvg Algorithm \cite{CELD}}\label{Alg1}
%  \KwIn{Set of participants $\mathcal{M}$ with size $|\mathcal{M}|$, local training dataset $\mathcal{D}_m$ with size $|\mathcal{D}_m|$ of participant $m\in\mathcal{M}$, number of global rounds $T$, local minibatch size $B$, number of local epochs $E$, and learning rate $\gamma$.}
%  \KwOut{Global model $\mathbf{w}_G^T$.}
%  Parameter server execution:\\
%  Initialize the global model $\mathbf{w}_G^0$ and distribute it to all participants\\
%  \For{$t=1$\quad\textbf{to}\quad$T$}
%  {
%     Choose a random subset $\mathcal{M}^t$ of $|\mathcal{M}^t|$ participants from the overall set of participants $\mathcal{M}$\\
%     \For{Each participant $j\in\mathcal{M}^t$ in parallel}
%     {
%        $\mathbf{w}_j^t\leftarrow$\textbf{LocalTraining}$(j, \mathbf{w}_G^{t-1})$
%     }
%     $\mathbf{w}_G^t\leftarrow\frac{1}{\sum_{j\in\mathcal{M}^t}|\mathcal{D}_j|}\sum_{j\in\mathcal{M}^t}|\mathcal{D}_j|\mathbf{w}_j^t$\\
%     Distribute $\mathbf{w}_G^t$ to all participants
%  }
%  Participant execution:\\
%  \textbf{LocalTraining}$(j, \mathbf{w})$\\
%  \For{$e=1$\quad\textbf{to}\quad$E$}
%  {
%     Divide the local dataset $\mathcal{D}_m$ to several minibatches with equal size $B$ that consists of a set $\mathcal{B}_m$ for each $m\in\mathcal{M}$\\
%     \For{each $b\in\mathcal{B}_m$}
%     {
%        $\mathbf{w}\leftarrow\mathbf{w}-\gamma\bigtriangleup f(\mathbf{w}; b)$ ($\bigtriangleup f$ is the gradient of $f$ on $b$)
%     }
%  }
%  Return $\mathbf{w}_G^T$
%\end{algorithm}
\subsection{Potential of FL for 6G Networks}
Federated learning has the potential to realize the ubiquitous intelligence envisioned by 6G networks while preserving the user privacy. Specifically, it can empower the edge intelligence even if the number of participants is in massive-scale and the collected data of different users is heterogeneous, which are supposed to be very common in 6G networks. Recent experimental results demonstrate that FL can converge to the optimum point, even when the number of participants is much larger than the average number of training samples in the dataset of each participant \cite{TFLS}. And both experimental and theoretical studies \cite{TFLS,OCFN} validate that FL supports model training among multiple users with non-identically distributed (non-IID) training datasets, whose convergence can be guaranteed in such a non-IID case. Moreover, since the information shared by each participant is about the learning model parameters rather than the original local data, the privacy of participants could be well protected. Nevertheless, the model update may also leak partial information of participants, while the security of FL could be further improved by exploiting more advanced security and encryption measures including differential privacy and secure aggregation.
\newcommand{\tabincell}[2]{\begin{tabular}{@{}#1@{}}#2\end{tabular}}

\begin{table*}[t]\footnotesize
%\vspace{-0.3cm}
%\setlength{\abovecaptionskip}{2pt}
%\setlength{\belowcaptionskip}{2pt}
\centering
\caption{A comparison of classical FL, general wireless FL, and AGIFL.}
\begin{tabular}{|c|c|c|c|}
  \hline
  \textbf{Works} & \textbf{Network Dimension} & \textbf{Application Scenarios} & \textbf{Key Design Factors} \\
  \hline
  \tabincell{c}{Classical FL\\ (\textit{e.g.}, \cite{CELD})} & Two-dimension & Training at terrestrial mobile devices & Training configuration  \\
  \hline
  \tabincell{c}{General wireless FL\\ (\textit{e.g.}, \cite{FLWN, AFLR})} & Two-dimension & Training in terrestrial wireless networks & Joint training configuration and wireless setting \\
  \hline
  Our proposed AGIFL & \textit{Three-dimension} & \textit{Training in air-ground integrated networks} & \tabincell{c}{Joint training configuration, wireless setting,\\ and \textit{air-ground network deployment}}\\
  \hline
\end{tabular}
\label{Novelty}
%\vspace{-0.4cm}
\end{table*}
\section{Air-Ground Integrated Federated Learning: Motivation, Overview, Vision, and Novelty}\label{Sec3}
In this section, we propose the framework of AGIFL, which features FL as a key enabler to empower the air-ground integrated networks intelligence. In the following, we first explain its motivation, then provide an overview of the AGIFL framework, and lastly present the vision.

\textbf{Motivation:} Recalling the potential of FL for 6G networks as in Sec.~\ref{Sec2}-B, the AGIFL framework is motivated by the following reasons. On one hand, as a critical part of the core potential architecture of future 6G networks, the AGINs are also expected to be with ubiquitous intelligence. More critically, it is envisioned that the required intelligence could be obtained locally at such edge networks for fast response and better data privacy protection, \textit{i.e.}, edge intelligence. On the other hand, as mentioned before, FL can enable multiple users with limited computation power and heterogeneous training dataset to collaboratively train an accurate learning model, without allowing the data leaving out of each user. And the FL architecture is highly flexible in the sense that it can take many forms as desired. Therefore, FL could perfectly solve the ubiquitous edge intelligence urgently needed by the AGINs.  % δÀ´ÍøÂçµÄÐèÇó£»ÁªºÏѧϰµÄ¸÷ÖÖÓÅÊÆ£¬½«Á½Õß½áºÏÕýºÃÄܽâ¾öÎÊÌâ

\textbf{Overview:} In general, the AGIFL framework includes different forms of FL between air and terrestrial networks, where the parameter server could be aerial node like a UAV or terrestrial BS according to different scenarios. First, exploiting the wide coverage and maneuverability of UAVs, when some terrestrial nodes such as mobile users need to collaboratively obtain a global learning model, a UAV can be deployed on-demand and act as the parameter server for these users. Secondly, a swarm of UAVs with collected local data as flying users may also exploit ML to execute complex tasks, where a terrestrial BS can be the parameter server to enable the FL among them. There are also other forms of FL in the proposed AGIFL framework, which will be introduced in detail in Sec.~\ref{Sec3}.

\textbf{Vision:} We envision that the AGIFL framework can efficiently address the challenge of edge intelligence generation in the AGINs. Specifically, based on the AGIFL framework, the edge users in the AGINs, \textit{e.g.}, mobile users, IoT devices, and UAVs, can support various on-device intelligent applications, independent of centralized data processing. Thus, the anticipated giant volume of data generated by billions of IoT devices at the network edge could be efficiently processed in a distributed way. In addition, we believe that, the privacy and security of the data distributed at different users can be protected in the AGIFL framework, which relieves the increasingly concern about the privacy issues.

\textbf{Novelty:} As shown in Table~\ref{Novelty}, our proposed AGIFL differs from the classical FL and general wireless FL, because it expands both the network dimension and application scenario, and considers more key design factors. First, the AGIFL could boost the intelligence for the AGINs in the three dimensions, while both classical FL \cite{CELD} and general wireless FL \cite{FLWN, AFLR} suit to the terrestrial networks in the two dimensions only. Secondly, compared to them, the AGIFL should consider the joint training configuration and wireless setting, under the flexible but complex air-ground network deployment. For example, the dynamically configurable location of aerial nodes could affect the energy consumption of both aerial and terrestrial training nodes, which results in the achieved learning performance, since there may be a strict energy budget of these nodes, especially in the flying aerial nodes. To summarize, the AGIFL is a new learning framework for realizing the ubiquitous intelligence within the AGINs, which brings a lot of research opportunities.

\section{Different Forms of AGIFL}\label{Sec4}
We now introduce all the forms of AGIFL one by one. The criterion for dividing the forms is set up by who the parameter server and training participants are, which is actually due to different intelligence needs. Note that we omit the ground to ground (G2G) federated learning in this work, since it is identical to the traditional FL.
\begin{figure*}[t]
%\vspace{-0.5cm}
\centering
%\subfloat[Varying $\tau$ (heterogeneous case)]{\includegraphics[height=1.1in]{s3_m12_heter_k}}\hfill
%\subfloat[Varying $B$ (heterogeneous case)]{\includegraphics[height=1.1in]{s3_m12_heter_budget}}\hfill
\subfloat[A2A federated learning]{\includegraphics[height=1.6in]{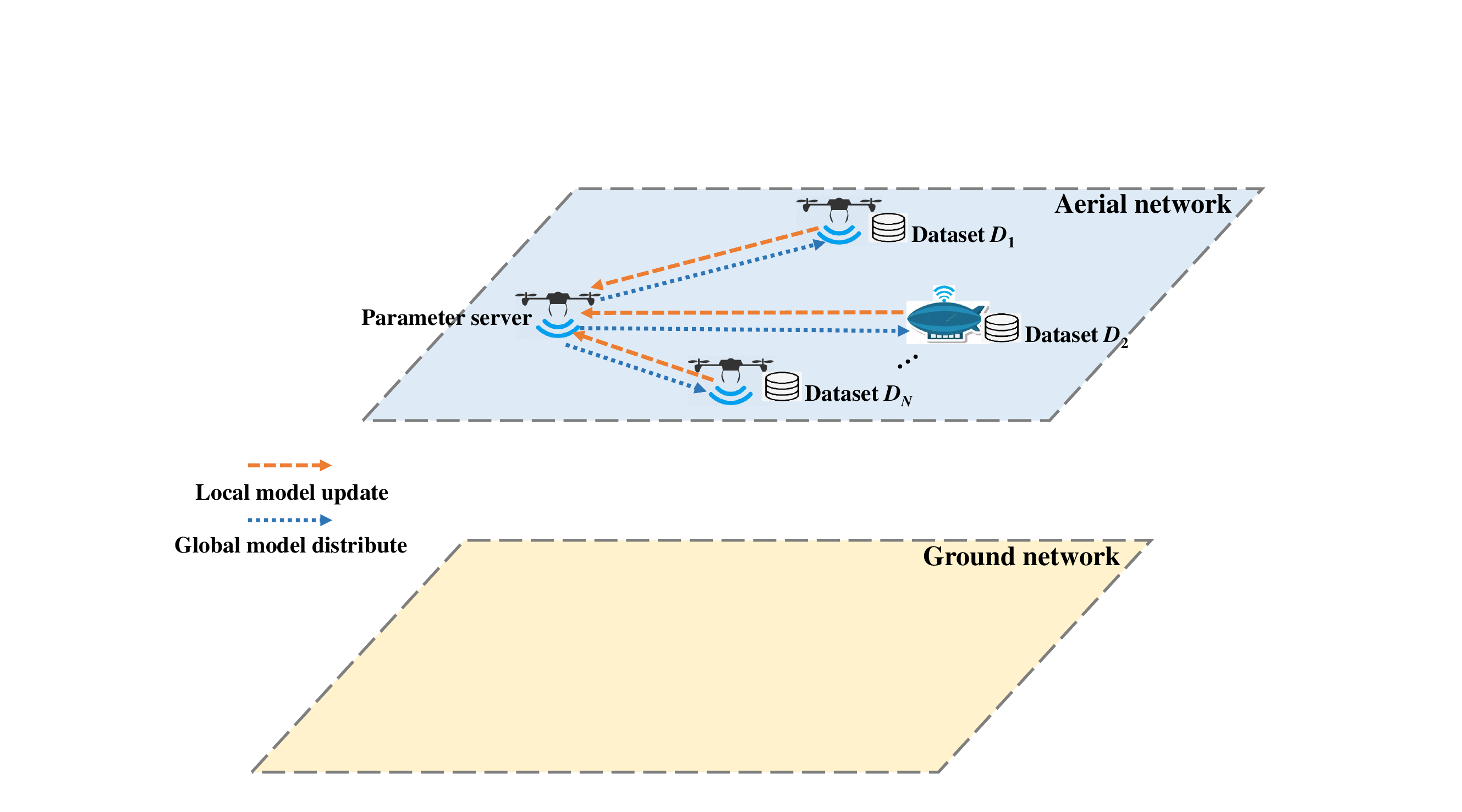}}\hspace{0.3cm}
\subfloat[G2A federated learning]{\includegraphics[height=1.6in]{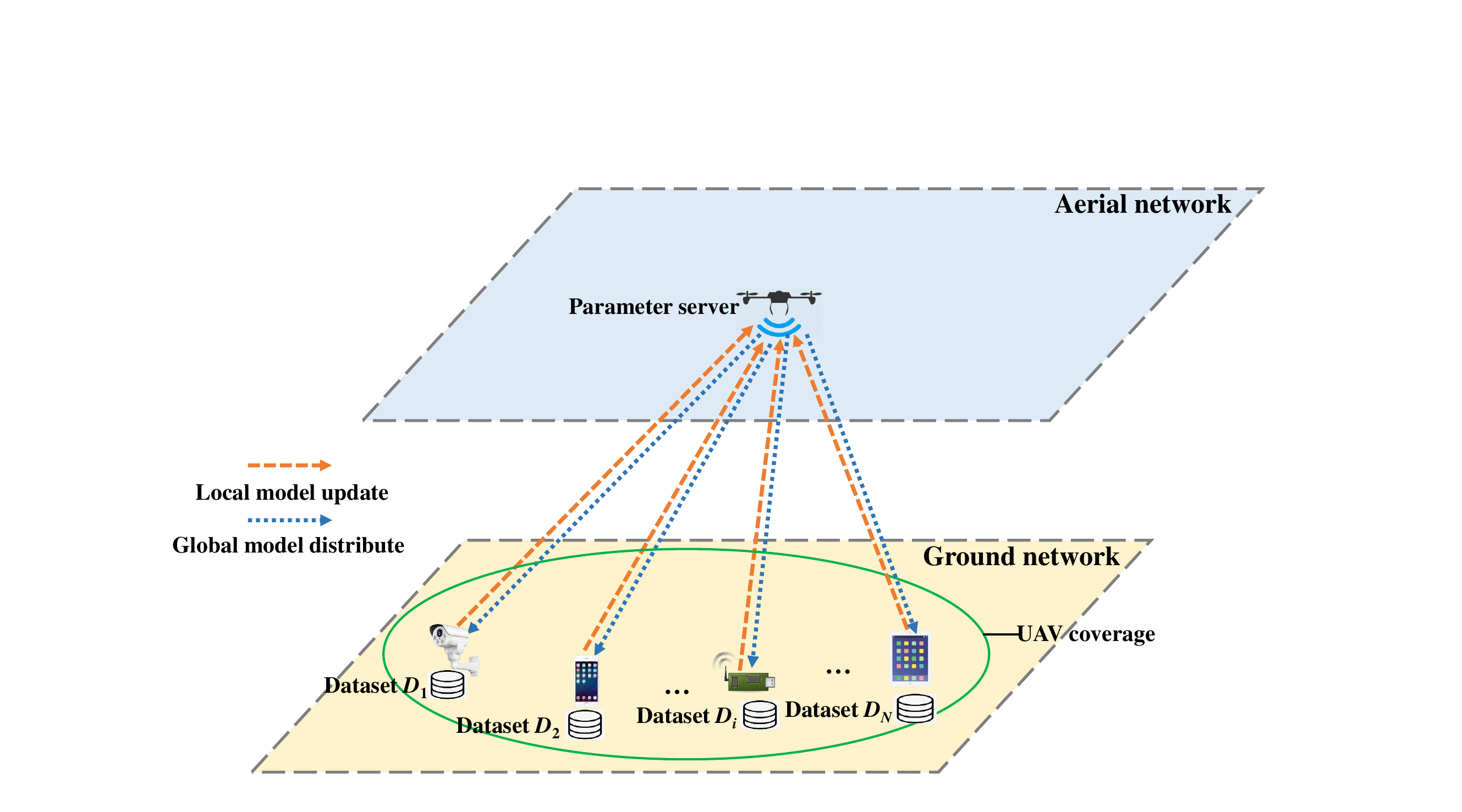}}\hfill\\
\subfloat[A2G federated learning]{\includegraphics[height=1.6in]{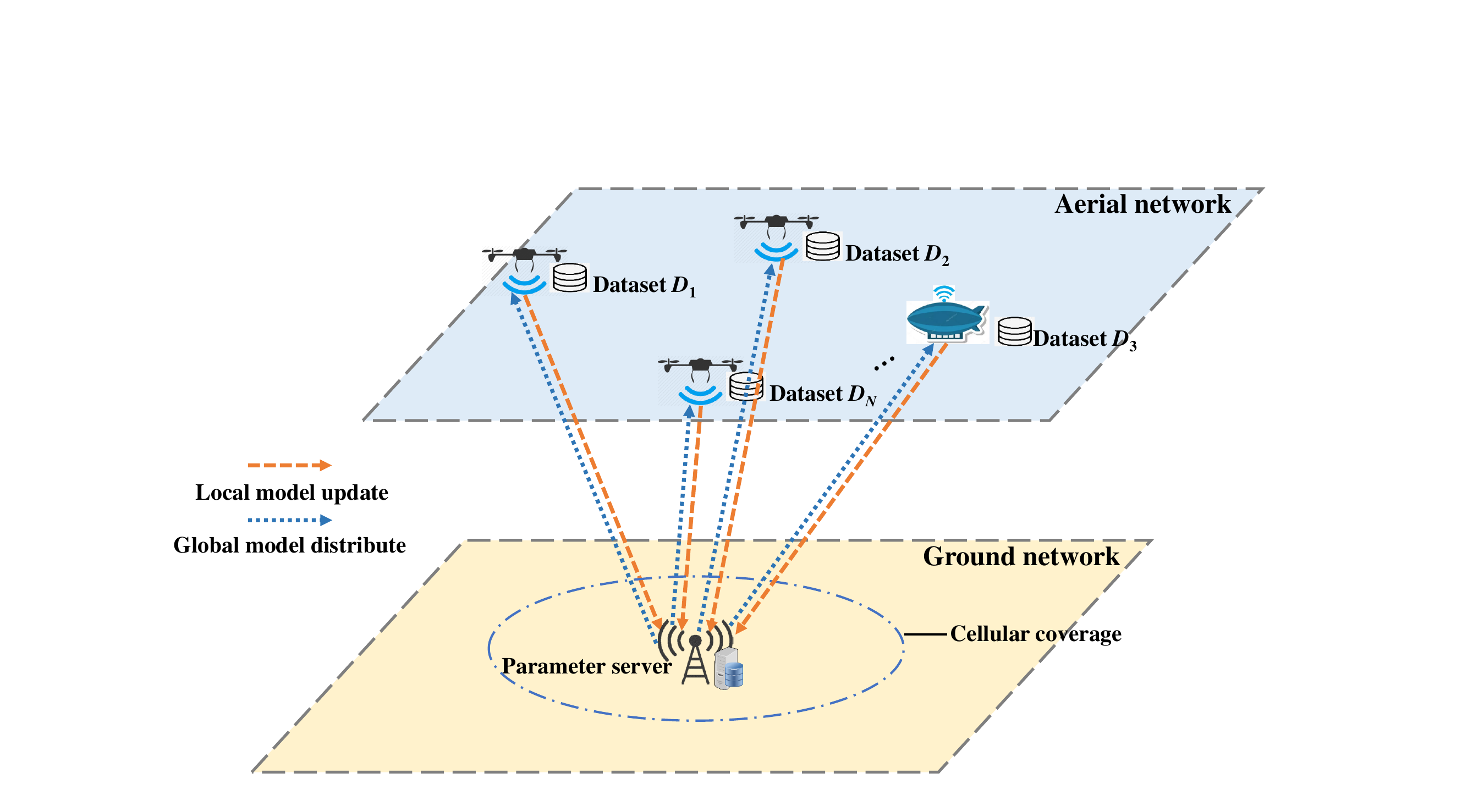}}\hspace{0.3cm}
\subfloat[Mixed federated learning]{\includegraphics[height=1.6in]{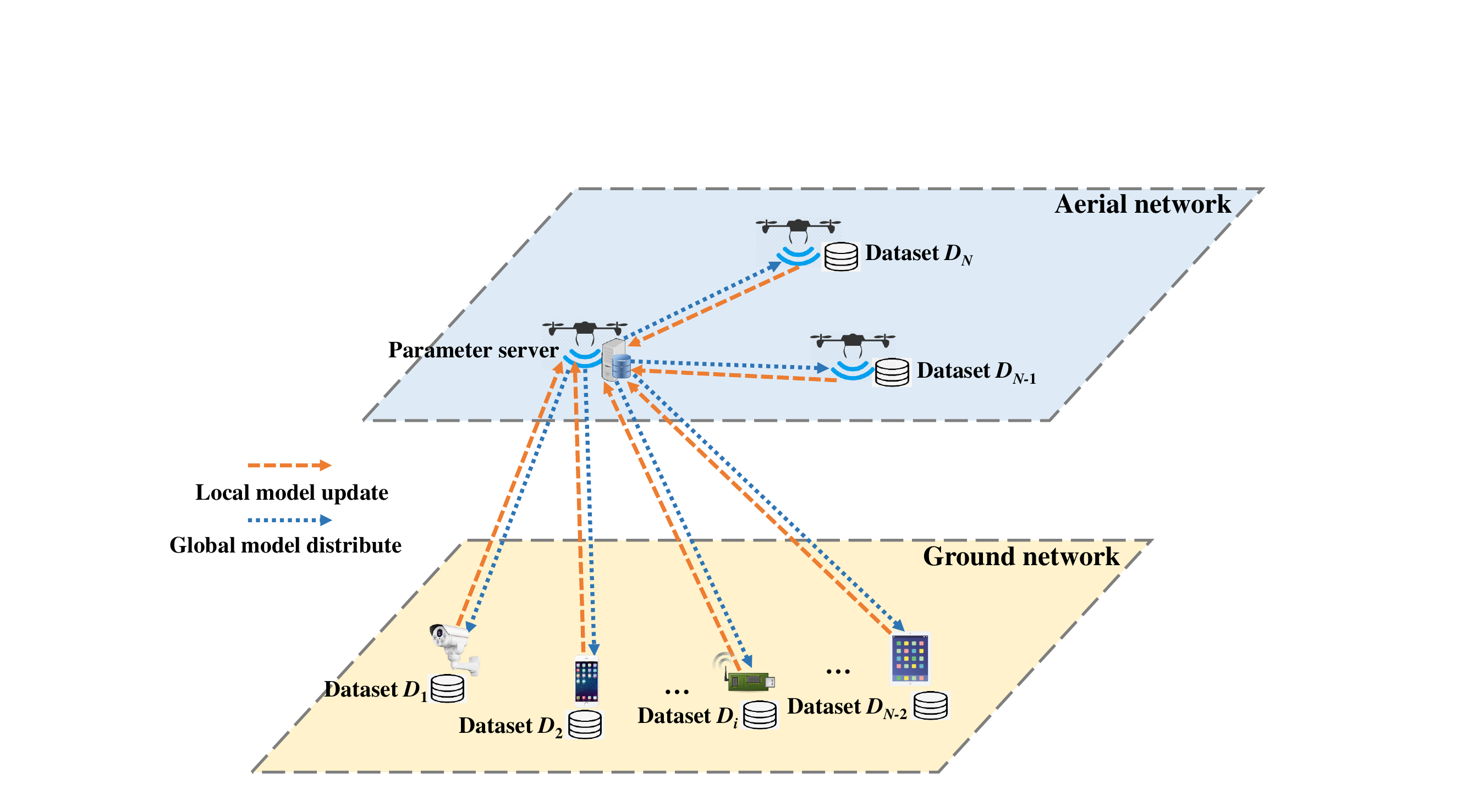}}\hfill
\caption{Different forms of federated learning in AGIFL.}
\label{AGIFL}
\vspace{-0.5cm}
\end{figure*}

%\begin{figure*}
%  \centering
%  % Requires \usepackage{graphicx}
%  \includegraphics[width=6in]{AGIFL}\\
%  \caption{Different forms of federated learning in AGIFL.}\label{AGIFL}
%\end{figure*}

\subsection{A2A Federated Learning}%Ê×ÏȽéÉÜÕûÌåÁ÷³Ì£¬Æä´Î½éÉܶÔӦDZÔÚÓ¦Óó¡¾°
First, we introduce the air to air (A2A) federated learning, where the training process happens within the aerial networks only. One typical example is the FL framework proposed in \cite{FLS}, which enables FL within a swarm of wirelessly connected UAVs. We generalize that framework as the A2A federated learning in the AGIFL, whose application scenario is shown in Fig.~\ref{AGIFL}~(a). In the A2A federated learning, one aerial node such as a leading UAV or unmanned airship plays a role of the parameter server, while several aerial nodes act as the training participants. Example applications of the A2A federated learning include various ML tasks execution in the sky, \textit{e.g.}, coordinated trajectory planning, cooperative target recognition, surveillance and monitoring for both military and civil use, with the help of a swarm of UAVs and unmanned airships, and etc.

\subsection{G2A Federated Learning}
Secondly, unlike the A2A federated learning within the aerial networks solely, we envision that the aerial platform could play a critical role in boosting ubiquitous edge intelligence for the terrestrial networks, which is termed as ground to air (G2A) federated learning in this work. To be specific, a UAV can be employed as the parameter server to meet the need of collaborative learning model training of multiple terrestrial nodes distributed over a large area, as shown in Fig.~\ref{AGIFL}~(b). One fascinating point of the G2A federated learning lies in that, combined with the wide coverage, the higher flexibility of aerial nodes' movement can be effectively utilized to enable the collaborative training of terrestrial networks on demand, without any ground infrastructure support. The G2A federated learning is very suitable for the Artificial Intelligence \& Internet of Things (AIoTs), where UAVs can be deployed on-demand to support the learning model training of massive number of IoT nodes. It is worth noting that this kind of FL has seldom been discussed in existing works.

\subsection{A2G Federated Learning}
Thirdly, different from the G2A federated learning, the edge intelligence of the aerial networks can also be boosted with the help of static or mobile terrestrial nodes, \textit{i.e.}, air to ground (A2G) federated learning. Specifically, as shown in Fig.~\ref{AGIFL}~(c), multiple aerial nodes with corresponding collected data train a ML model in a federated way, when a terrestrial node such as ground BS is employed as the parameter server. For example, for the traffic prediction over a target region in an Intelligent Transport System (ITS), several UAVs covering different sub-regions can be employed to first collect traffic flow from road side units (RSUs) or capture real-time traffic data over highways, and then return to the corresponding UAV bases for collaborative model training, under the guidance of a terrestrial model owner as the parameter server \cite{TFLU}. Compared to the A2A federated learning, the A2G federated learning might enable more powerful intelligence for the aerial networks, since UAVs with limited energy budget need not be as the parameter server besides the training participants, and the A2G communication link may be better than the A2A link. We believe that in future 6G networks, the edge intelligence could be obtained in such a mixed federated way.

\subsection{Mixed Federated Learning}
Lastly, we also anticipate that the AGIFL can work in a highly flexible manner in the AGINs, known as mixed federated learning, \textit{i.e.}, both aerial and terrestrial nodes could be as the training participants and parameter server as required. As Fig.~\ref{AGIFL}~(d) shows, a number of terrestrial devices and UAVs, both with private collected raw data, aims to train a FL model, where a UAV covering those nodes acts as the aggregated parameter server. It is worthy noting that a terrestrial node such as a ground BS can also be the parameter server, and a UAV can be the model parameter forwarding node, by exploiting its wide coverage as well as flexible deployment.

\section{Technical Challenges}\label{Sec5}
The features of AGINs including three-dimensional mobility, frequent inevitable air-ground interactions, and relatively constrained resources such as communication, computation, and energy, together with unique learning requirements, result in difficulties in learning-oriented network design and optimization. Note that the aforementioned features distinguish the proposed AGIFL from both classical FL and general wireless FL, since those features can affect the training performance of FL in terms of training latency and energy consumption, or even training accuracy considering limited energy budget of aerial/terrestrial nodes. In this following, we pose three main technical challenges in detail in the AGIFL.
\subsection{On-Demand Network Deployment for AGIFL}
The aerial layer in the AGINs is dynamically configurable by nature, \textit{i.e.}, the aerial nodes such as UAVs and unmanned airships are generally deployed on-demand. Thus, it is inevitable to optimize the deployment of the aerial nodes to cater to the FL model training, \textit{e.g.}, how to determine the optimal location of the aerial node as the parameter server in the G2A and mixed FL. However, it is challenging to optimally deploy the aerial nodes in the AGIFL in the following aspects.

\textbf{Learning-oriented deployment:} different from maximizing the quality-of-service/experience (QoS/QoE) in traditional AGINs, maximizing the learning performance such as maximizing training accuracy and minimizing training latency is the main optimization objective in the AGIFL, which obviously alters the optimal deployment decision of the aerial nodes to achieve that goal. It is also different from the deployment of traditional terrestrial edge computing servers \cite{ECEL}.

\textbf{Long-term plan of deployment:} since the model training in FL always lasts for multiple training rounds, the deployment decision of the aerial nodes should account for the long-term training profit, rather than the short-term profit within one single round, which is hard to optimize due to the complexity of multi-stage training.

\textbf{On-demand movement trajectory:} in some dynamic scenarios, \textit{e.g.}, the training participants in the G2A FL are smart vehicles, the location of the aerial node as the parameter server should not be static and thus its movement trajectory should be determined, which is extremely challenging to tackle. Specifically, following the energy consumption model in \cite{FDAT}, the consumed energy of each mobile participant in each training round is affected by the location of the parameter server UAV, because the transmission rate of the UAV-ground wireless channel is also affected by the distance between the UAV and each participant. Similarly, the latency of the G2A FL per round might be different due to that varying transmission rate, which needs further investigation.
\subsection{Learning-Oriented Resource Allocation and Training Configuration}
In the AGIFL, the current AGIN is built up for boosting the learning performance of the model training. This calls for the learning-oriented joint resource allocation and training configuration, which differs from both the design of AGINs \cite{AGIME} and FL optimization for general wireless networks \cite{FLWN,AFLR}. Specifically, the optimization of the current AGINs is learning-oriented, while that of the general AGINs is QoS/QoE-oriented. In addition, although there emerge several recent studies about how to optimize the performances of FL in wireless networks \cite{FLWN,AFLR}, they do not consider the on-demand deployment of aerial nodes and apply to the FL within the terrestrial networks only.

There exist two main difficulties in solving the above problem. First, the problem involves multiple variables to be jointly optimized, \textit{i.e.}, the location of the aerial node, the allocation of different resources in terms of communication and computation, and the decision of training parameters such as number of training rounds, number of chosen participants per round, number of the local updates, and mini-batch size. It makes the above problem a natural mixed-integral problem (MIP) and is hard to solve in general. Secondly, as it needs multiple air-ground interactions over the inherent unreliable wireless channel, how to design robust training strategies to guarantee the overall learning performance against the unreliable data transmission in the AGIFL is hard. Specifically, the unreliable data transmission may lead to the loss of some updated local or global model parameter in any training round, which obviously slows down the training process. Even worse, the loss of the model update is uncertain owing to the uncertain wireless channel loss, thereby increasing the difficulty of designing robust training strategies.
%\subsection{Energy Efficient Training Strategies for AGIFL}
%Pursuing high energy efficiency is critical for realizing its potential benefits of the AGIFL, since the model training may cost much energy consumption and the energy budget of the aerial nodes as well as terrestrial nodes including IoT nodes is limited. Recalling the training process of FL, the training participants undertake the most computation work and consume some energy in transmitting local model updates, while the parameter server consumes less energy because the simple aggregation of local model updates consumes little energy, besides the almost same energy consumption of broadcasting global model update. In the AGIFL, putting more weight on which part in optimizing the energy consumption depends on the exact form. For instance, in the G2A FL, it is more reasonable to optimize the energy consumption of the terrestrial nodes to maximize the learning performance or guarantee the minimum learning performance. In contrast, in the A2G FL, we should be more concerned with the energy efficiency of the aerial nodes, as any aerial node will spend much energy in model training, communicating, and flying. As seen in the above, the energy efficiency problem varies with different forms of the AGIFL. Additionally, the energy model of the aerial nodes is more complex than that of the terrestrial ones. In a word, it is very challenging to design energy efficient strategies for the AGIFL.
\subsection{Management of Control and Data in AGIFL}
Note that there exist many critical decisions to be made in the AGIFL, \textit{e.g.}, the decisions about the aforementioned on-demand network deployment, resource allocation and training configuration, and training strategies. Thus, an interesting but challenging question is \textit{who is in charge of those decisions in the AGIFL}? Furthermore, to make correct decisions, it needs to collect enough data about the network environment information, so \textit{how the data is collected and processed}? Some fixed BSs with powerful communication and computation abilities could be a good candidate, which may incur additional overhead including extra communications between aerial/terrestrial nodes and those BSs. In addition, to enable the cooperation among multiple UAVs in the AGIFL, it is inevitable to share some information such as their locations, remaining battery powers, movement trajectories, and other parameters, whose update frequency should be intelligently controlled to maintain a tradeoff between the FL performance and incurred extra overhead. Another issue about data in the AGIFL is how both aerial and terrestrial nodes can obtain enough labeled training datasets for the federated training, especially for those UAVs. We believe that, initially, those nodes could individually obtain preliminary data and label the data based on the local or partial observation of some well-known phenomenons. And they could obtain more labeled data with the help of terrestrial users. In general, the process of obtaining enough appropriate training datasets is also worth further studies.
\section{Case Study: G2A Federated Learning}
We conduct a case study to evaluate the performance of the proposed AGIFL framework by numerical simulations. To be specific, we focus on the G2A FL where a UAV is employed as the parameter server for the FL of multiple terrestrial nodes. And we mainly evaluate the effect of different UAV's hovering location deployment schemes on the learning performance and UAV's energy consumption.

%\begin{figure*}[t]
%\centering
%\subfloat[Energy consumption \textit{vs.} training rounds]{\includegraphics[height=2.5in]{Energy}}\hfill
%\subfloat[Accuracy \textit{vs.} UAV's energy budget]{\includegraphics[height=2.5in]{Accuracy}}
%\caption{Performance comparison of different UAV deployment schemes.}
%\label{Performance}
%%\vspace{-0.5cm}
%\end{figure*}

\subsection{Settings}
We consider a UAV-assisted network where a rotary-wing UAV with the ability to hover over a set $\mathcal{U}$ of $U=100$ terrestrial users, each with a local training dataset. Those users are randomly distributed in a 100$\times$100 m$^2$ area. Employing the UAV as the parameter server, these users collaboratively train a learning model for inference, where the model training requires interactions between the UAV and users within multiple rounds. In each round, the UAV aggregates a global model and distributes it to the users, and each user then updates its local model by its own dataset and sends it to the UAV. Following \cite{CELD}, we choose only a random fraction $\theta=0.1$ of users for model update at the beginning of each round. And the learning rate is fixed at 0.002 and the number of local epochs is set as 5 with the mini-batch size 10. In this simulation, for the training task, we consider the image classification using convolutional neural network (CNN) on the classical MNIST dataset \cite{CELD}. The whole training dataset is partitioned in those UEs, each of which receives the same examples (\textit{i.e.}, IID setting). As for the FL algorithm, we use the celebrated FedAverage algorithm \cite{CELD}. In the future, we would also like to investigate the effects of non-IID dataset setting and other FL algorithms such as FedProx in the proposed AGIFL.

For the UAV, we suppose its transmission power and propulsion power equal to 10 mW and 100 W, respectively \cite{EEUC}. The overall energy consumption of the UAV mainly consists of the hovering energy consumption (the product of propulsion power and hovering time), and the energy consumption for aggregated model broadcast (the product of transmission power and broadcast time). For the users, we assume the computation capacity of user $u\in\mathcal{U}$ and number of CPU cycles to execute one bit of the sample data are chosen randomly from the interval [1.8 GHz, 2.0 GHz] and fixed at 10, respectively. As to the communication between user $u\in\mathcal{U}$ and the UAV, the uplink data rate is given by $r_u=B_u\log_2\left(1+\frac{\alpha_0 p_u}{\sigma^2(H^2+R^2_u)}\right)$, where $B_u$ is the spectrum bandwidth allocated to user $u$, $p_u$ is the data transmission power of user $u$, $\alpha_0$ is the channel power gain at the reference distance 1 m, $H$ is the fixed flying height of the UAV, $R_u$ is the horizontal distance between the UAV and user $u$, and $\sigma^2$ is the noise power. Similar to \cite{EEUC}, we set the values of the above parameters as follows: $B_u=\frac{1}{\theta U}B$ ($B=1$ MHz), $p_u=100$ mW, $\alpha_0$ = -50 dB, $\sigma^2=$ -90 dBm, and $H=$ 100 m. Similarly, the downlink data rate can be easily calculated, except that $B_u$ and $p_u$ should be 1 MHz and 10 mW for the UAV, respectively.

Assume the coordinate of the $u$-th user is known as $[x_u, y_u]$, for $u\in\mathcal{U}$, and we need to determine the coordinate of the UAV $[X, Y, H]$, where the flying height $H$ is fixed at 100 m as previously mentioned. We consider two different deployment schemes of the UAV's hovering location as follows:
\begin{itemize}
      \item \textbf{Min\_SumDist}, it finds a pair $(X^*, Y^*)$ such that the sum of the distances between the UAV and each user is minimized, \textit{i.e.}, $(X^*, Y^*)\in\arg\min_{(X, Y)} \sum_{u=1}^U d_u$, where $d_u:= \sqrt{(X-x_u)^2 + (Y-y_u)^2+H^2}$.
%      \item \textbf{Min\_MaxDist}, it finds a pair $(X^*, Y^*)$ such that the maximum distance between the UAV and each user is minimized, \textit{i.e.,}, $(X^*, Y^*)\in\arg\min_{(X, Y)} ~\max~\{d_u\}_{u=1}^U$.
      \item \textbf{Random}, it randomly chooses the hovering location of the UAV within the current area.
\end{itemize}Note that in the former scheme, the optimization problem is convex, which thus can be easily solved by existing convex optimization techniques. In the simulation, all results are averaged over 20 repeated simulation instances.

\subsection{Result Analysis}
\begin{figure}
  \centering
  % Requires \usepackage{graphicx}
  \includegraphics[width=3.2in]{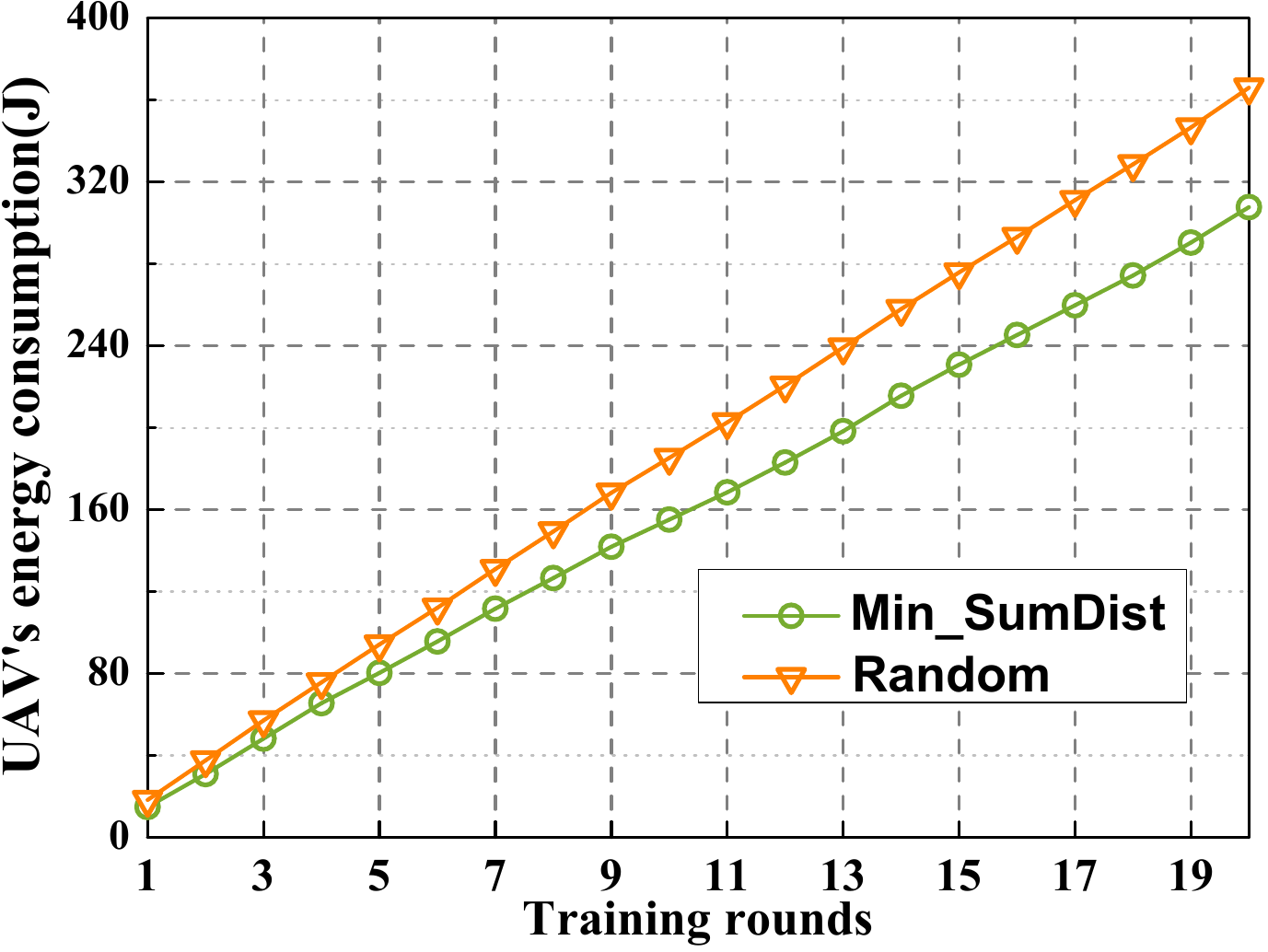}\\
  \caption{Energy consumption \textit{vs.} training rounds.}\label{Performance_E}
\end{figure}

\begin{figure}
  \centering
  % Requires \usepackage{graphicx}
  \includegraphics[width=3.2in]{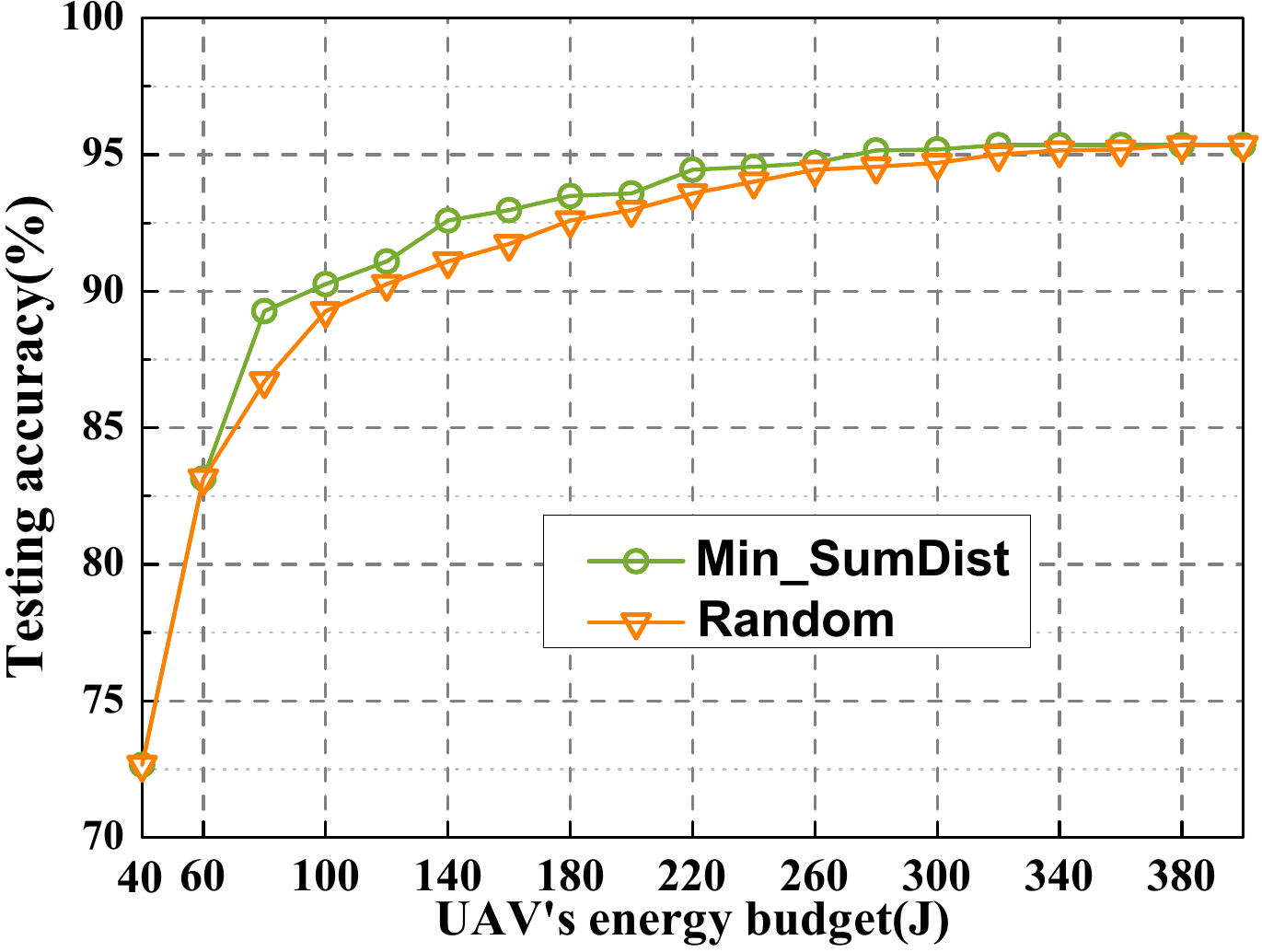}\\
  \caption{Testing accuracy \textit{vs.} UAV's energy budget.}\label{Performance_A}
\end{figure}

\begin{figure}
  \centering
  % Requires \usepackage{graphicx}
  \includegraphics[width=3.2in]{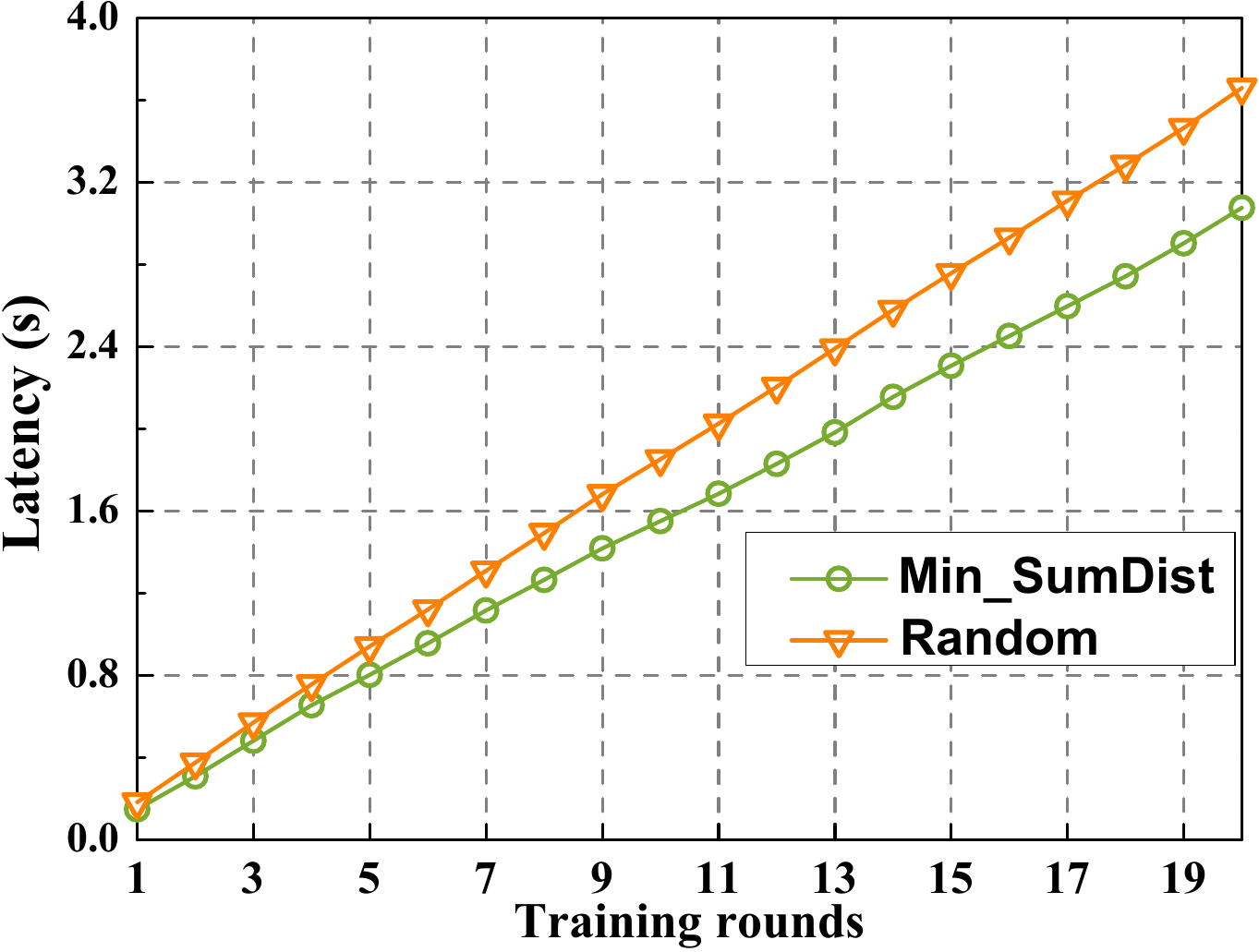}\\
  \caption{Training latency \textit{vs.} training rounds.}\label{Performance_L}
\end{figure}

Fig.~\ref{Performance_E}, Fig.~\ref{Performance_A}, and Fig.~\ref{Performance_L} present the performance comparison of the two deployment schemes in terms of the UAV's energy consumption, testing accuracy, and overall training latency, respectively, with the variation of the number of training rounds and UAV's energy budget. First, according to Fig.~\ref{Performance_E}, the UAV's overall energy consumption under the Min\_SumDist scheme is always less than that under the Random scheme, whose gap increases with the number of training rounds. This implies that the hovering location of the UAV should be carefully optimized, because the UAV usually has a strict energy budget in practical. Secondly, as shown in Fig.~\ref{Performance_A}, when the UAV's energy budget is increased, the maximum testing accuracy is also higher, because the training could be run more rounds, with the increased energy budget. And when the UAV's energy budget is about more than 340 J, the testing accuracies achieved by both schemes become very close, since when there is enough energy, it approaches to the maximum accuracy by the current dataset and training parameters. Furthermore, we find that the Min\_SumDist scheme achieves higher accuracy than the Random scheme, since the former consumes less energy than the latter with the same number of training rounds. Thirdly, similar to Fig.~\ref{Performance_E}, Fig.~\ref{Performance_L} shows that, the overall training latency of both deployment schemes almost increases linearly with the number of training rounds due to the identical training and communication strategies at each round, while the Min\_SumDist always achieves smaller latency than the Random. To sum up, the deployment of the UAV as well its inborn energy constraint has a great impact on the performance of the AGIFL.

%Intelligence is an important characteristic of 6G networks. Combined with AI, 6G networks can learn to achieve self-configuration, self-optimization, self-organization, and self-healing. (The intelligence will play a more prominent role in the network, going beyond classification and prediction tasks which are being considered for 5G systems).
%
%The 6G vision for a data-driven society, enabled by near-instant and unlimited wireless connectivity, will require addressing complex, heterogeneous and often conflicting design requirements.

\section{Future Research Directions}
Despite its great potential, the study on the AGIFL is still in its infancy, where many key research issues need to be addressed. In this section, we discuss several potential research directions for future study.

\textbf{Joint Optimization of UAVs' Locations, Resource Allocation, and Training Parameters in AGIFL:} Motivated by the challenges stated in Sec.~\ref{Sec5}, a critical but challenging research question is how to jointly optimize UAVs' locations, resource allocation, and training parameters, to boost the learning performance. There are numerous studies about how to jointly optimize UAVs' locations/movement trajectories and resource allocation in existing works, whose solutions however cannot be applied to that problem, since it involves various additional training variables and is a brand-new problem. Furthermore, the problem form as well as its related constraints is different in different forms of the AGIFL, so there may not exist a general solution framework to it.

\textbf{Block Chain-Based Secure Model Parameters Exchange for AGIFL:} Although FL enables multiple devices to collaboratively train a shared learning model without exchanging their private local data, there still exists a potential risk of information leakage, as the model parameter exchange may also be used to derive some private information by the malicious participants. Block chain is a recently emerging distributed ledger technology where a network of participants reach agreements on shared data, independent of a central trusted authority. Intuitively, it is worthy to study how to exploit block chain in the AGIFL for the data security.

\textbf{Intelligent Collaboration among Multiple Aerial Nodes for AGIFL:} It usually needs multiple aerial nodes including UAVs and unmanned airships for model training in the AGIFL. For example, some aerial nodes may act as the model update forwarders for some terrestrial participants in the mixed FL. Thanks to the aggregation of model update by weighted averaging in FL, an aerial node as the model update forwarder could aggregate the received local updates in advance and then send the partial aggregated model update to the parameter server. Therefore, it is interesting to investigate how many aerial nodes are needed to cover a number of terrestrial participants and how to intelligently collaborate them.

\textbf{Decentralized AGIFL without a Central Parameter Server:} The current AGIFL employs the most widely used FL strategy, which relies on a central node as the aggregated parameter server. The robustness as well as the scalability is thus a big issue, \textit{e.g.}, the proposed AGIFL cannot work if the wireless links to the central parameter server fail, or the central parameter node such as a UAV is attacked or out of battery. It is more attractive to develop a fully decentralized AGIFL framework without a central parameter server to adapt the dynamic environment of the AGINs, \textit{e.g.}, each terrestrial node or UAV only exchanges local model parameters with its neighboring node or UAV for model aggregation. The decentralized AGIFL is more flexible and robust, but may introduce much overhead because of a large amount of model parameter exchange among the training participants.%This especially suits for the A2A FL, since such a centralized framework may be prone to failure due to the breakdown of some aerial node as the parameter server.

\section{Conclusion}
In this article, we have proposed the AGIFL framework to empower the edge intelligence to realize the required ubiquitous intelligence for future 6G networks. Specifically, the AGIFL utilizes FL to enable all nodes in the AGINs to collaboratively train a learning model, with the help of flexible controllable deployment of aerial nodes. We have introduced the basic concept of FL and its potential, described the overall framework as well as its different forms, discussed several main technical challenges, conducted a case study of AGIFL, and highlighted some promising future research topics.

\section*{Acknowledgement}
This work is supported in part by the National Key R\&D Program of China No. 2018YFB1800800, in part by the National Natural Science Foundation of China under Grant No. 61931011, Grant No. 62072303, Grant No. 61872178, Grant No. 61801505, in part by National Postdoctoral Program for Innovative Talents of China No. BX20190202, in part by the Open Project Program of the Key Laboratory of Dynamic Cognitive System of Electromagnetic Spectrum Space No. KF20202105, in part by the Natural Science Foundation of Jiangsu Province under Grant No. BK20181251, in part by the Fundamental Research Funds for the Central Universities under Grant 14380059, in part by the funding from Hong Kong RGC Research Impact Fund (RIF) with the Project No. R5034-18 and the General Research Fund of the Research Grants Council of Hong Kong (PolyU 152221/19E).

\bibliographystyle{IEEEtran}
%\bibliography{IEEEabrv}

\footnotesize
\begin{thebibliography}{26}

\bibitem{AI6G}
H. Yang, A. Alphones, Z. Xiong, D. Niyato, J. Zhao, and K. Wu, ``Artificial intelligence-enabled intelligent 6G networks,'' \textit{IEEE Network}, vol. 34, no. 6, pp. 272-280, 2020.

%\bibitem{R6G}
%K. B. Letaief, W. Chen, Y. Shi, J. Zhang, and Y. A. Zhang, ``The roadmap to 6G: AI empowered wireless networks,'' \textit{IEEE Commun. Mag.}, vol. 57, no. 8, pp. 84-90, 2019.

\bibitem{V6G}
W. Saad, M. Bennis, and M. Chen, ``A vision of 6G wireless systems: Applications, trends, technologies, and open research problems,'' \textit{IEEE Network}, vol. 34, no. 3, pp. 134-142, 2020.

%\bibitem{FLSI}
%J. Konecn\'{y}, H. B., McMahan, F. X., Yu, P. Richt\'{a}rik, A. T. Suresh, and D. Bacon, ``Federated learning: Strategies for improving communication efficiency,'' \textit{CoRR}, abs/1610.05492, 2016.

\bibitem{CELD}
B. McMahan, E. Moore, D. Ramage, S. Hampson, and B. A. Arcas, ``Communication-efficient learning of deep networks from decentralized data,'' in \textit{Proc. AISTATS}, pp. 1273-1282, 2017.

\bibitem{FDAT}
X. Diao, J. Zheng, Y. Cai, Y. Wu, and A. Anpalagan, ``Fair data allocation and trajectory optimization for UAV-assisted mobile edge computing,'' \textit{IEEE Communications Letters}, vol. 23, no. 12, pp. 2357-2361, 2019.

\bibitem{FLS}
T. Zeng, O. Semiari, M. Mozaffari, M. Chen, W. Saad, and M. Bennis, ``Federated learning in the sky: Joint power allocation and scheduling with UAV swarms,'' \textit{Proc. IEEE ICC}, pp. 1-6, 2020.

\bibitem{CEMU}
H. Shiri, J. Park, and M. Bennis, ``Communication-efficient massive uav online path control: Federated learning meets mean-field game theory,'' \textit{IEEE Transactions on Communications}, vol. 68, no. 11, pp. 6840-6857, 2020.

\bibitem{TFLU}
W. Y. B. Lim, J. Huang, Z. Xiong, J. Kang, D. Niyato, X.-S. Hua, C. Leung, and C. Miao, ``Towards federated learning in uav-enabled Internet of vehicles: A multi-dimensional contact-matching approach,'' \textit{IEEE Transactions on Intelligent Transportation Systems}, DOI: 10.1109/TITS.2021.3056341, 2021.

\bibitem{FLUE}
B. Brik, A. Ksentini, and M. Bouaziz, ``Federated learning for UAVs-enabled wireless networks: Use cases, challenges, and open problems,'' \textit{IEEE ACCESS}, 8: 53841-53849, 2020.

\bibitem{TFLS}
K. Bonawitz \textit{et al.}, ``Towards federated learning at scale: System design,'' \textit{Proc. MLSys}, 2019.

\bibitem{OCFN}
X. Li, K. Huang, W. Yang, S. Wang, and Z. Zhang, ``On the convergence of fedavg on non-iid data,'' \textit{Proc. ICLR}, 2020.

\bibitem{ECEL}
C. Xiang, Z. Zhang, Y. Qu, D. Lu, X. Fan, P. Yang, and F. Wu, ``Edge computing-empowered large-scale traffic data recovery leveraging low-rank theory,'' \textit{IEEE Transactions on Network Science and Engineering}, vol. 7, no. 4, pp. 2205-2218, 2020.

\bibitem{AGIME}
N. Cheng, W. Xu, W. Shi, Y. Zhou, N. Lu, H. Zhou, and X. Shen, ``Air-ground integrated mobile edge networks: Architecture, challenges, and opportunities,'' \textit{IEEE Communications Magazine}, vol. 56, no. 8, pp. 26-32, 2018.

\bibitem{FLWN}
N. H. Tran, W. Bao, A. Zomaya, M. N. H. Nguren, and C. S. Hong, ``Federated learning over wireless networks: Optimization model design and analysis,'' \textit{in Proc. IEEE INFOCOM}, pp. 1387-1395, 2019.

\bibitem{AFLR}
S. Wang, T. Tuor, T. Salonidis, K. K. Leung, C. Makaya, T. He, and K. Chan, ``Adaptive federated learning in resource constrained edge computing systems,'' \textit{IEEE Journal on Selected Areas in Communications}, vol. 37, no. 6, pp. 1205-1221, 2019.

\bibitem{EEUC}
Y. Zeng and R. Zhang, ``Energy-efficient uav communication with trajectory optimization,'' \textit{IEEE Transactions on Wireless Communications}, vol. 16, no. 6, pp. 3747-3760, 2017.

\end{thebibliography}

\begin{IEEEbiography}[{\includegraphics[width=0.8in,height=1in,clip,keepaspectratio]{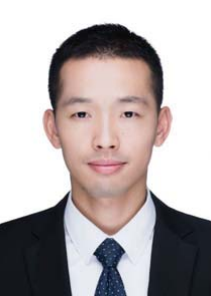}}]{Yuben Qu} received the Ph.D degree in Computer Science and Technology from Nanjing Institute of Communication Engineering, China in 2016. He is a research assistant in Nanjing University of Aeronautics and Astronautics, China. His current research interests include mobile edge computing, air-ground integrated networks, and federated learning.
\end{IEEEbiography}

\begin{IEEEbiography}[{\includegraphics[width=0.8in,height=1in,clip,keepaspectratio]{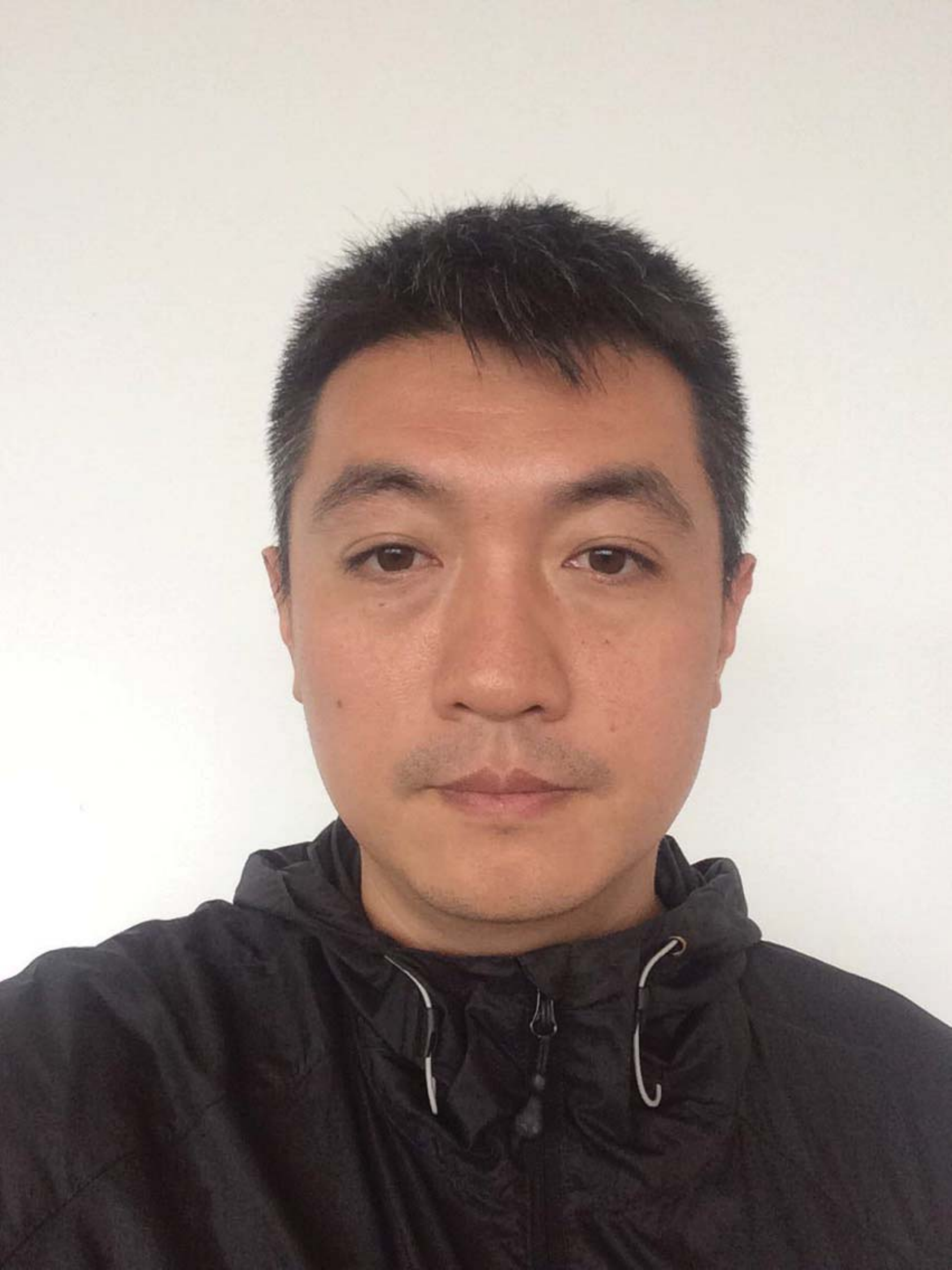}}]{Chao Dong} received his Ph.D degree in Communication Engineering from PLA University of Science and Technology, China, in 2007. He is now a full professor with the College of Electronic and Information Engineering, Nanjing University of Aeronautics and Astronautics, China. His current research interests include D2D, UAVs networking, and anti-jamming.
\end{IEEEbiography}

\begin{IEEEbiography}[{\includegraphics[width=0.8in,height=1in,clip,keepaspectratio]{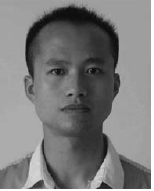}}]{Jianchao Zheng} received the Ph.D. degree in communications and information systems from PLA University of Science and Technology, China in 2016. He is currently an Associate Professor at the National Innovation Institute of Defense Technology. His research interests focus on green communications and computing networks, game theory, and optimization techniques.
\end{IEEEbiography}

\begin{IEEEbiography}[{\includegraphics [width=0.8in,height=1in,clip,keepaspectratio] {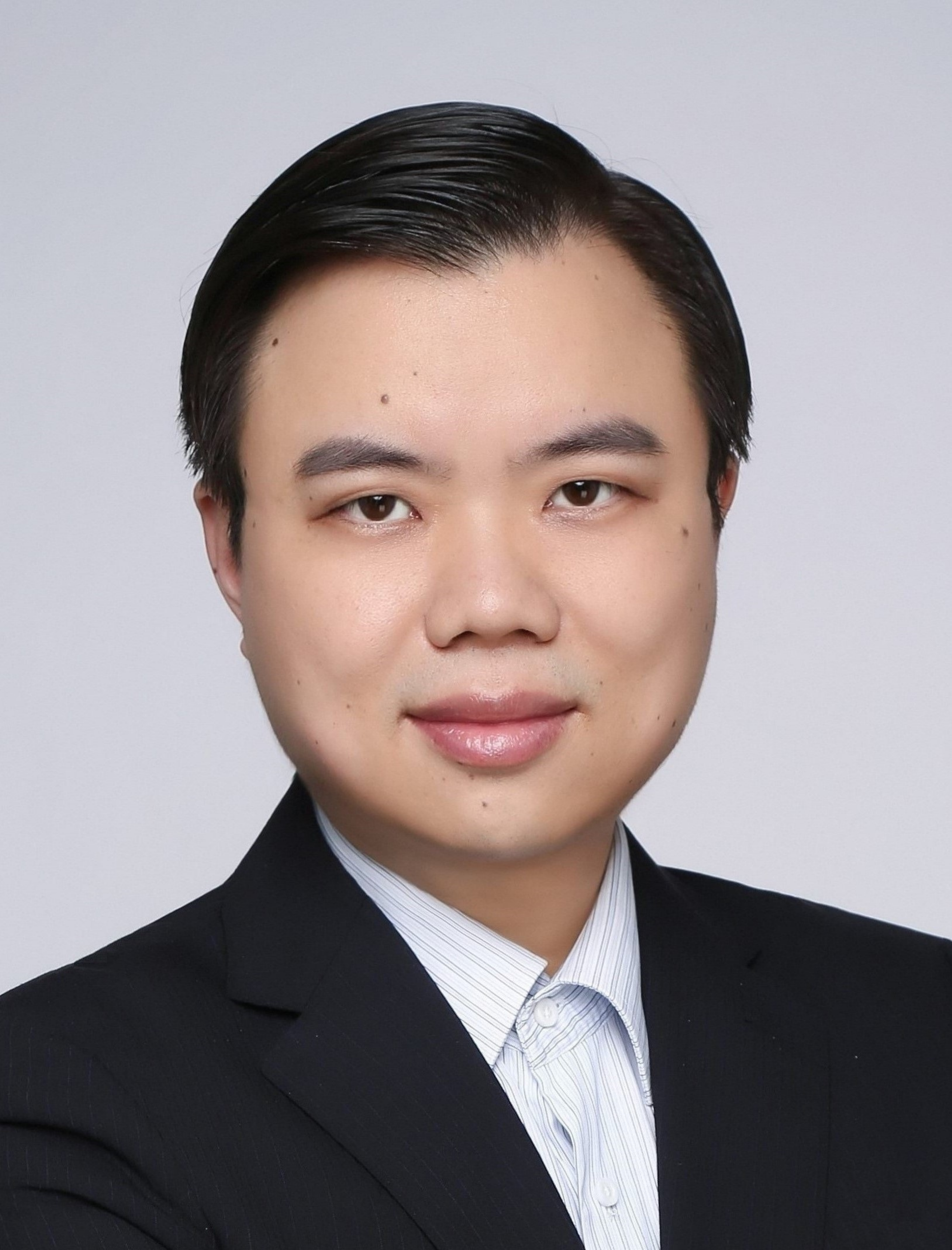}}]{Haipeng Dai} received the Ph.D. degree in Computer Science and Technology in Nanjing University, China in 2014. He is currently an associate professor in the Department of Computer Science and Technology in Nanjing University. His research interests are mainly in the areas of wireless charging, mobile computing, and data mining.
\end{IEEEbiography}

\begin{IEEEbiography}[{\includegraphics[width=0.8in,height=1in,clip,keepaspectratio]{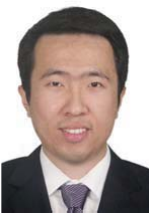}}]{Fan Wu} is a professor in the Department of Computer Science and Engineering, Shanghai Jiao Tong University. He received his Ph.D. in Computer Science and Engineering from the State University of New York at Buffalo in 2009. His research interests include wireless networking and computing, game theory, and privacy preservation.
\end{IEEEbiography}

\begin{IEEEbiography}[{\includegraphics[width=0.8in,height=1in,clip,keepaspectratio]{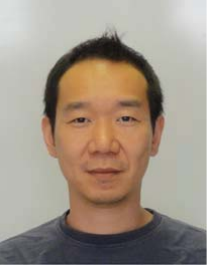}}]{Song Guo} is a Full Professor at Department of Computing, The Hong Kong Polytechnic University. His research interests are mainly in the areas of big data, cloud computing, mobile computing, and distributed systems with over 450 papers published in major conferences and journals. He is a fellow of the IEEE.
\end{IEEEbiography}

\begin{IEEEbiography}[{\includegraphics[width=1in,height=3in,clip,keepaspectratio]{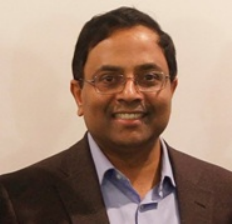}}]{Alagan Anpalagan} is a professor at the Department of Electrical and Computer Engineering, Ryerson University, Canada. He has co-authored five edited books and co-chaired several conferences. He is a fellow of the IET and EIC. He received his PhD degree in Electrical Engineering from the University of Toronto, Canada.
\end{IEEEbiography}

\end{document}